\documentclass[aip, apl, superscriptaddress, reprint]{revtex4-1}

\usepackage{siunitx}
\usepackage{chemformula}
\usepackage{graphicx}
\usepackage{hyperref}
\usepackage{verbatim}

\newcommand{\mr}[1]{\mathrm{#1}}
\newcommand{\ea}[0]{\textit{et al.}}
\newcommand{\cf}[0]{cf.\ }
\newcommand{\ie}[0]{i.e.\ }

\newcommand{\fref}[1]{Fig.~\ref{fig:#1}}

\newcommand{\Cref}[1]{Chapter~\ref{chap:#1}}
\newcommand{\cref}[1]{Ch.~\ref{chap:#1}}

\renewcommand{\vec}[1]{{\mathrm{\mathbf{#1}}}}

\begin{document}

\title{Evolution of the Spin Hall Magnetoresistance in \ch{Cr2O3}/Pt bilayers close to the N\'e{}el temperature}

\author{Richard Schlitz}
\email{richard.schlitz@tu-dresden.de}
\affiliation{Institut f{\"u}r Festk{\"o}rper- und Materialphysik, Technische Universit{\"a}t Dresden, 01062 Dresden, Germany}
\affiliation{Center for Transport and Devices of Emergent Materials, Technische Universit{\"a}t Dresden, 01062 Dresden, Germany}
\author{Tobias Kosub}
\affiliation{Helmholtz-Zentrum Dresden-Rossendorf e.V., Institute of Ion Beam Physics and Materials Research, 01328 Dresden, Germany}
\author{Andy Thomas}
\affiliation{Leibniz Institute for Solid State and Materials Research Dresden (IFW Dresden), Institute for Metallic Materials, 01069 Dresden, Germany}
\affiliation{Center for Transport and Devices of Emergent Materials, Technische Universit{\"a}t Dresden, 01062 Dresden, Germany}
\author{Savio Fabretti}
\affiliation{Institut f{\"u}r Festk{\"o}rper- und Materialphysik, Technische Universit{\"a}t Dresden, 01062 Dresden, Germany}
\affiliation{Center for Transport and Devices of Emergent Materials, Technische Universit{\"a}t Dresden, 01062 Dresden, Germany}
\author{Kornelius Nielsch}
\affiliation{Leibniz Institute for Solid State and Materials Research Dresden (IFW Dresden), Institute for Metallic Materials, 01069 Dresden, Germany}
\affiliation{Technische Universit{\"a}t Dresden, Institute of Materials Science, 01062 Dresden, Germany}
\author{Denys Makarov}
\affiliation{Helmholtz-Zentrum Dresden-Rossendorf e.V., Institute of Ion Beam Physics and Materials Research, 01328 Dresden, Germany}
\author{Sebastian T. B. Goennenwein}
\affiliation{Institut f{\"u}r Festk{\"o}rper- und Materialphysik, Technische Universit{\"a}t Dresden, 01062 Dresden, Germany}
\affiliation{Center for Transport and Devices of Emergent Materials, Technische Universit{\"a}t Dresden, 01062 Dresden, Germany}
\date{\today}

\begin{abstract}
    We study the evolution of magnetoresistance with temperature in thin film bilayers consisting of platinum and the antiferromagnet \ch{Cr2O3} with its easy axis out of the plane. We vary the temperature from \SIrange{20}{60}{\celsius},  close to the N\'e{}el temperature of \ch{Cr2O3} of approximately \SI{37}{\celsius}. The magnetoresistive response is recorded during rotations of the external magnetic field in three mutually orthogonal planes. A large magnetoresistance having a symmetry consistent with a positive spin Hall magnetoresistance is observed in the paramagnetic phase of the \ch{Cr2O3}, which however vanishes when cooling to below the N\'e{}el temperature. Comparing to analogous experiments in a \ch{Gd3Ga5O12}/Pt heterostructure, we conclude that a paramagnetic field induced magnetization in the insulator is not sufficient to explain the observed magnetoresistance. We speculate that the type of magnetic moments at the interface qualitatively impacts the spin angular momentum transfer, with the $3d$ moments of \ch{Cr} sinking angular momentum much more efficiently as compared to the more localized $4f$ moments of \ch{Gd}.
\end{abstract}

\maketitle

The spin Hall magnetoresistance (SMR), \cite{ChenSMR, AltiSMR, NakayamaSMR} arising from the combined action of the spin Hall\cite{Hirsch:1999} and inverse spin Hall effect is a powerful tool to monitor the magnetization direction in a ferrimagnetic insulator (FMI)/normal metal(NM) hybrid structure.\cite{AltiSMR, NakayamaSMR} Recent publications furthermore established that the SMR also allows to probe the surface magnetization \cite{Isasa:2014}, and to resolve exotic magnetic phases such as spin canting\cite{Ganzhorn:2016} and helical magnetic order.\cite{Aqeel:2016} In particular, magnetic phase diagrams could be reconstructed from the SMR response \cite{Ganzhorn:2016}, since the SMR is  sensitive to the sublattice magnetization orientations.\cite{Gomonay:2014} Last but not least, increasing interest in antiferromagnetic spintronics\cite{Wadley:2016, Marti:2014, Jungwirth:2016, Zelezny:2014} led to theoretical investigations\cite{Wang:2017} and experimental studies of the spin Hall magnetoresistance in antiferromagnetic insulator (AFMI)/NM heterostructures.\cite{Hoogeboom:2017, Gepraegs:2017, Baldrati:2017, Ji:2017} 

In a FMI/NM bilayer, it often is sufficient to consider the (net) magnetization $\vec{m}$ in the FMI. The SMR can then be expressed as\cite{ChenSMR} 
\begin{equation}
	R_\mr{long} = R_0 - \Delta R \, m_t^2 = R_0 - \Delta R \sin^2(\angle(\vec{m, t})))
	\label{eq:SMR}
\end{equation}
where $\Delta R > 0$ denotes the change in resistance arising as a function of the orientation of the magnetization unit vector $\vec{m}$. $m_t$ is the projection of the magnetization on the direction $\vec{t}$ perpendicular to the current direction $\vec{j}$ as well as the surface normal $\vec{n}$. $\vec{j}$, $\vec{t}$ and $\vec{n}$ are an orthonormal set of unit vectors. According to Eq.\,\ref{eq:SMR}, the fingerprint of the SMR effect thus is a maximum resistance for $\vec{m}\parallel\vec{j}$ viz.\,$\vec{m}\parallel\vec{n}$, and a smaller resistance otherwise. Additionally, since $R_\mr{long}$ is proportional to $\sin^2(\angle(\vec{m}, \vec{t}))$ the SMR is symmetric upon magnetization reversal. 

Interestingly, SMR experiments in AFMI/NM heterostructures point to a more complex  behavior.\cite{Hoogeboom:2017, Gepraegs:2017, Baldrati:2017, Ji:2017} However, most of the experiments performed to date were carried out in the easy plane AFM NiO, where multiple domains with equivalent energy coexist and can exhibit different magnetoresistance responses. Since the magnetic configuration of an easy axis AFM is much simpler, SMR experiments in such materials appear desirable. \ch{Cr2O3} is a prototypical easy axis AFM, i.e., this material features one uniaxial anisotropy easy axis, and it was recently established for applications in antiferromagnetic spintronics.\cite{Kosub:2015, Kosub:2016} Furthermore, the N\'e{}el temperature  $T_{\mathrm{N}}\approx \SI{37}{\celsius}$ of bulk \ch{Cr2O3} \cite{He:2010} is close to room temperature, allowing for magnetotransport experiments across the AFM/paramagnetic phase transition using a simple thermoelectric cooler/heater. 

Thus, we have fabricated thin film heterostructures consisting of \ch{Cr2O3} and platinum, and measured the evolution of their magnetoresistive response  across $T_{\mathrm{N}}$. Interestingly, we observe a clear SMR signal for $T>T_{\mathrm{N}}$, while the magnetoresistive response below $T_{\mathrm{N}}$ is very small. Our experiments thus show that the SMR in AFMI/NM heterostructures can be used to monitor the N\'e{}el transition. Moreover, the evolution of the magnetoresistance with external field orientation above the N\'e{}el temperature is consistent with Eq.\,\ref{eq:SMR}, as expected for an SMR signal originating from the field induced paramagnetic polarization of the AFMI layer. This paramagnetic polarization will always follow the external magnetic field, giving rise to a finite positive magnetoresistance as observed in FMI/Pt heterostructures.\cite{Aqeel:2015,AltiSMR} Since we here rotate the external magnetic field in three mutually orthogonal planes, we can unambiguously identify the signature of the SMR in our AFMI/NM heterostructures.

The samples were prepared as follows: A \SI{250}{\nm} \ch{Cr2O3} layer was grown onto c-cut \ch{Al2O3} substrates using reactive evaporation of \ch{Cr} from a Knudsen cell in \SI{1e-5}{\milli\bar} molecular oxygen gas. The substrate temperature was set to \SI{700}{\celsius} initially and lowered to \SI{500}{\celsius} after the growth of the first few monolayers.
Directly after growth, the \ch{Cr2O3} layer was annealed in vacuum at \SI{750}{\celsius}. This leads to an easy axis of the antiferromagnetic anisotropy in the out of plane orientation. Subsequently, a $2-\SI{3}{\nm}$ platinum film was sputtered in-situ at \SI{100}{\celsius}, resulting in the layer sequence schematically depicted in \fref{rot}(a). Note that \ch{Cr} is expected at the \ch{Cr2O3}/Pt interface in our samples, which is favorable compared to oxygen terminated interfaces due to the magnetoelectric properties of \ch{Cr2O3}.\cite{Belashchenko:2010, Ning:2011}
For further details about the sample structure please refer to our previous work.\cite{Kosub:2015, Kosub:2016, KosubPhD:2016} We then patterned the \ch{Cr2O3}/Pt bilayers into Hall-bars (cf.\ \fref{rot}(b)) using optical lithography and inverse sputtering with \ch{Ar}-ions. The contacts used for the voltage measurement are separated by $l = \SI{300}{\um}$. The width of the Hall-bar is $w = \SI{40}{\um}$. Finally, the samples were glued onto chip carriers, electrically contacted via wedge bonding and mounted into a magnetotransport setup to characterize the magnetoresistive response. The setup features a cylindrical Halbach array\cite{HalbachArray} generating a constant magnetic flux density $\mu_0 H = \SI{1}{\tesla}$ perpendicular to the array's cylindrical axis. We control the sample temperature with a thermoelectric cooler attached to the sample inserts, enabling us to vary the temperature in the range of \SIrange{20}{60}{\celsius}. The measured temperatures are detected in close vicinity of the sample position with a Pt100 resistance thermometer.

\begin{figure}[th]
    \includegraphics{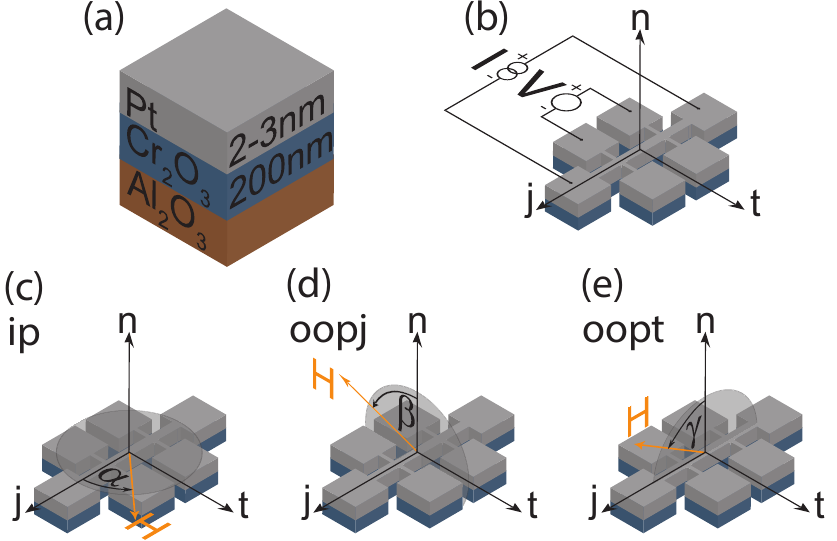}
    \caption{\label{fig:rot}Panel (a) displays the layer sequence after deposition. The coordinate system spanned by the current direction $\vec{j}$, the surface normal $\vec{n}$ and the transverse direction $\vec{t}$ as well as the patterned Hall-bar are shown in panel (b).  A constant current of $I = \SI{180}{\uA}$ is applied along the Hall-bar and the voltage drop is simultaneously recorded in a four point measurement scheme. The definitions of the three mutually orthogonal magnetic field rotation planes with their angles $\alpha$, $\beta$ and $\gamma$ are depicted in panels (c), (d) and (e), respectively.}
\end{figure}

To obtain the magnetoresistive response, we drive a current of $I = \SI{180}{\uA}$ along the Hall-bar with a Keithley 2450 sourcemeter. The voltage drop is simultaneously detected by a Keithley 2182 nanovoltmeter. To increase the measurement sensitivity and to remove thermoelectric contributions, we employ a current reversal technique.\cite{Goennenwein:2015} In the magnetotransport experiments, the Halbach array and, thus, the magnetic field is rotated around the cylindrical axis. By mounting the sample in three different sample inserts, we define three orthogonal rotation planes of the magnetic field. For the first two, the magnetic field is rotated around the surface normal $\vec{n}$ (\fref{rot}(c), ip) and around the direction of current flow $\vec{j}$ (\fref{rot}(d), oopj), respectively. In the third case, the magnetic field is rotated around the transverse direction to get a complete set of (orthogonal) rotation planes. 

The obtained magnetoresistance 
\begin{equation*}
    \frac{R(\alpha, \beta, \gamma)}{\min(R)} -1
\end{equation*}
of the \ch{Cr2O3}/Pt sample is shown in \fref{admr} for three different temperatures. The data is corrected to remove linear drifts. Here, the minimum resistance is $\min(R) = $\ \SIlist{526.5;539.8;547.5}{\ohm} for $T=$\ \SIlist{21.6; 37.6; 47}{\celsius} for the in-plane data, respectively.

\begin{figure}[th]
    \includegraphics{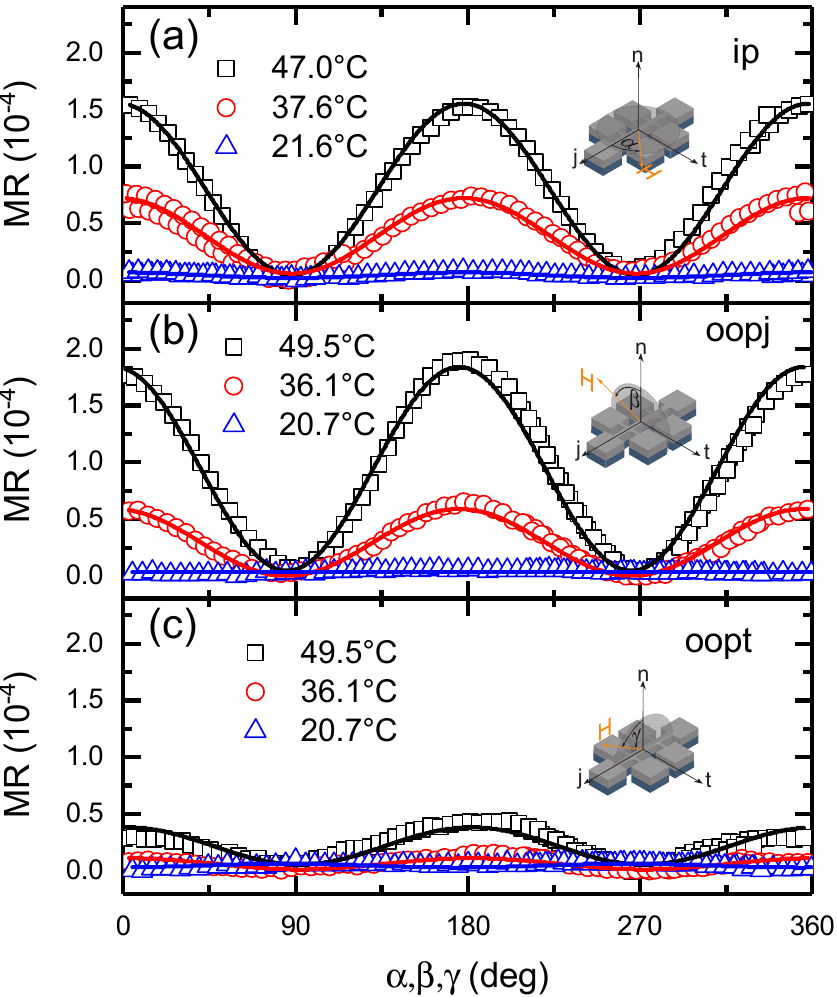}
    \caption{\label{fig:admr} The magnetoresistance obtained during rotations of the magnetic flux density $\mu_0 H = \SI{1}{\tesla}$ in ip, oopj and oopt configuration for three temperatures are shown in (a), (b) and (c), respectively. The black squares correspond to measurements above the N\'e{}el temperature. The data represented by the red circles were measured close to the N\'e{}el temperature, while the data shown as blue triangles were recorded just below. $\sin^2(\alpha, \beta, \gamma)$ fits, shown as lines, were performed to extract the amplitude of the signals. A linear slope was subtracted to remove drifts. 
    }
\end{figure}

We start the discussion with the data obtained during the ip and oopj rotation of the magnetic field (\cf \fref{admr}(a,b, black squares). Here, a $\sin^2(\alpha)$ modulation is evident at $T=\SI{47}{\celsius}>T_N$. When the temperature is approaching $T_N$, the amplitude of the resistance modulation decreases. For $T=\SI{20}{\celsius}$, the resistance modulation as a function of magnetic field orientation vanishes within our experimental resolution, as one naively would expect considering that the external magnetic field does not affect the AFM spin structure. Below this temperature range, condensation of ambient moisture jeopardizes reliable data taking in the setup used here. 

It is important to note that a small $\sin^2(\gamma)$ modulation is also observed for the oopt rotation, leading to the conclusion that in addition to a strong SMR, another effect is present in our structures. It could be attributed to a magnetic proximity effect, as suggested by the field invariant anomalous Hall contribution evident in our as well as similar samples.\cite{Kosub:2015, Kosub:2016} However, the finite resistance modulation in the oopt rotation could also reflect the crystallinity of the Pt film, giving rise to additional terms in the resistivity tensor due to symmetry as already observed for the anisotropic magnetoresistance.\cite{Limmer:2006}

By fitting a $\sin^2(\alpha, \beta, \gamma)$ to the data, we can extract the relative magnetoresistance amplitudes for all temperatures and rotation geometries. The obtained results are summarized in \fref{SMRofT}. To determine the exact N\'e{}el temperature in our heterostructures, we carried out zero-offset anomalous Hall magnetometry measurements.\cite{Kosub:2015} The resulting $R_\mr{inv}$ for positive and negative cooling field is shown in the upper panel of \fref{SMRofT} as black and red symbols, respectively, and yields a N\'e{}el temperature of $T_N = \SI{37}{\celsius}$ as indicated by the dashed orange line. 
This is in excellent agreement with the bulk \ch{Cr2O3} N\'e{}el temperature\cite{He:2010} of $T_N = \SI{37}{\celsius}$, and matches the temperature region where the SMR response changes. 
The SMR thus reflects the change in antiferromagnetic to paramagnetic order in the insulator. 
Interestingly, the change in SMR appears to be smeared out in a much broader temperature range as compared to the zero-offset anomalous Hall magnetometry data. 
\begin{figure}[th]
    \includegraphics{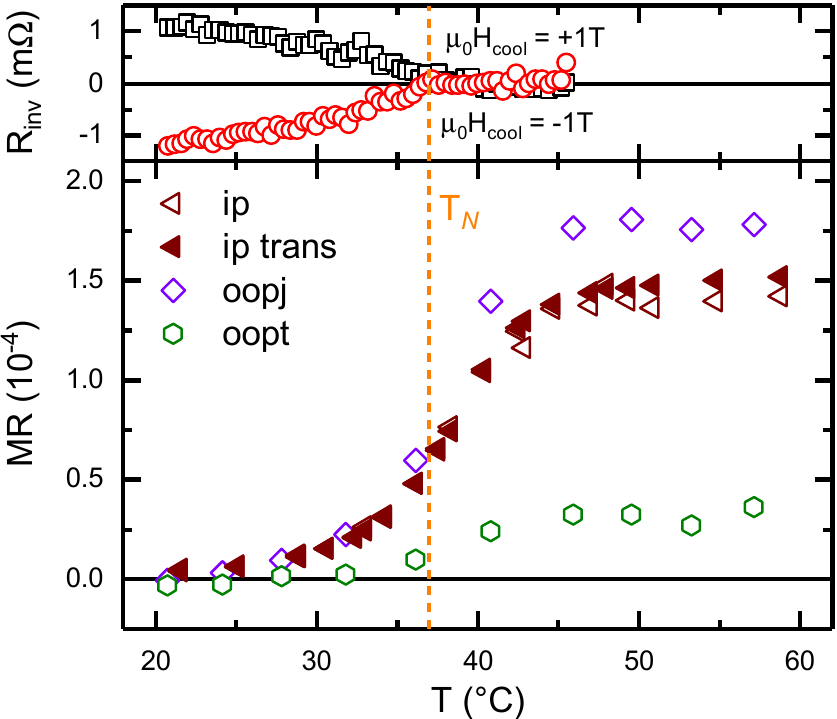}
	\caption{\label{fig:SMRofT}The upper panel shows the magnetic field invarant contribution $R_\mr{inv}$ to the Hall resisitivity acquired by zero-offset anomalous Hall magnetometry for positive (black symbols) and negative cooling field (red symbols)\cite{Kosub:2015}, vanishing at $T_N$. The magnetoresistance $\Delta R/R_0$ of the sample obtained from $\sin^2(\alpha, \beta, \gamma)$ fits to the angle dependent data for all three rotation planes and temperatures are summarized in the lower panel. The N\'e{}el temperature is indicated by the orange dashed line.
   A vanishing MR is observed below $T_N$ for all three rotation configurations. When increasing the temperature, the MR increases until $T\sim\SI{45}{\celsius}$ where it seems to saturate.} 
\end{figure}

This suggests that the SMR probes a different subset of magnetic moments, since the SMR increases further with increasing temperature even in the paramagnetic phase, saturating at a level of $\Delta R/R>\SI{1e-4}{}$ around $T\sim\SI{45}{\celsius}$. A non-vanishing MR above the Curie temperature was inferred by Aqeel \ea\ in bilayers of the ferrimagnetic insulator \ch{Cu2OSeO3} and Pt via ip rotations only, and attributed to a field induced paramagnetic magnetization acting as sink for the spin accumulation at the Pt interface.\cite{Aqeel:2015} However, the MR discussed by Aqeel \ea\ amounts to about $\Delta R_\mr{trans} \lesssim \SI{0.07}{\milli\ohm}$ in the paramagnetic phase. Taking the thickness of their platinum film $t_\mr{Pt}=\SI{5}{\nm}$ and a specific resistance of Pt of $\rho \sim \SI{200}{\nano\ohm\meter}$, this translates to a magnitude of the MR of $\Delta \rho_\mr{trans}/\rho \lesssim \SI{2e-6}{}$. Therefore, the magnetoresistance observed by Aqeel \ea\ is roughly two orders of magnitude smaller than the paramangetic SMR reported here. Furthermore, no saturation was observed by these authors even at \SI{200}{\kelvin} above the Curie temperature\cite{Aqeel:2015}, in contrast to the behavior in our samples evident from \fref{SMRofT}. We also would like to stress again that the MR we observe in three orthogonal rotation planes has the symmetry characteristic of SMR, well above $T_N$. 

The saturation value of $\left(\Delta R / R\right)_\mr{oopj} = \SI{1.8e-4}{}$ we observe here is smaller by roughly a factor of 7 when compared to the best YIG/Pt heterostructures\cite{AltiSMR} with comparable platinum thickness. This indicates a good interfacial quality of our heterostructures. Furthermore, in terms of angular momentum sinking capability across the interface to a metal, \ch{Cr2O3} in the paramagnetic phase thus still appears to be comparable to a good ferrimagnet. In addition, the SMR signal persists well above $T_{\mathrm{N}}$, with an apparent saturation or $T$-independent level in the range $\SI{45}{\celsius} <T<\SI{60}{\celsius}$.

We tentatively attribute the large SMR observed in the paramagnetic phase to a transfer of angular momentum from the Pt onto paramagnetic \ch{Cr} moments in \ch{Cr2O3}. However, control experiments in a \ch{Gd3Ga5O12}/Pt heterostructure with a platinum thickness of \SI{3}{\nm} reveal a vanishing MR, \ie $\Delta R/R \le \SI{1e-6}{}$, around room temperature. Thus, the presence of a finite, magnetic field induced paramagnetic magnetization apparently is not sufficient for the occurrence of a large SMR. In addition, the increase of the SMR magnitude with increasing $T$ even above the ordering temperature is counter-intuitive for a paramagnet. 

More quantitatively, the magnetic susceptibility of \ch{Gd3Ga5O12} (GGG) at room temperature is $\chi_{GGG}\approx \SI{6.8e-3}{}$ according to both a Curie-Weiss susceptibility calculation and experiments.\cite{Kim:2015} In comparison, Foner reported the susceptibility of \ch{Cr2O3} to be $\chi_{\ch{Cr2O3}}\approx \SI{1.6e-3}{}$ at room temperature\cite{Foner:1963}, such that the magnetic field induced magnetization in GGG should be about a factor 4 larger than in \ch{Cr2O3} in the temperature range of interest here. In spite of this comparable magnetization, the SMR observed in the respective heterostructures of \ch{Cr2O3} and \ch{Gd3Ga5O12} with Pt differs by more than 2 orders of magnitude. This could be due to the fact that the magnetic moments in GGG are the $4f$ moments of Gd, which are strongly localized and thus might not couple well to the spin accumulation in Pt. In other words, the effective mixing conductance in GGG/Pt could be much smaller than in \ch{Cr2O3}/Pt, since for the latter the coupling is mediated by the $3d$ moments on Cr. This would be consistent with the observations by Aqeel \ea\ for \ch{Cu2OSeO3}/Pt, where $3d$ moments are responsible for the magnetism. However, more comprehensive SMR experiments in different paramagnetic insulator/Pt heterostructures will be needed in the future to fully clarify the microscopic nature of the paramagnetic SMR effect.

In summary, we measured the magnetoresistive response in \ch{Cr2O3}/Pt heterostructures for different temperatures close to the N\'e{}el temperature. Comparing the MR in three mutually orthogonal rotation planes, we find a signal consistent with a positive SMR with a magnitude $\lesssim \SI{2e-4}{}$ several $\SI{10}{K}$ above the N\'e{}el temperature, which we attribute to a field-induced paramagnetic magnetization in the \ch{Cr2O3}. Furthermore, our experiments reveal that the mechanism leading to a finite SMR in the paramagnetic phase can not be attributed solely to the field induced magnetization, as no SMR was observed in GGG/Pt heterostructures, hinting at the microscopic mechanisms involved in the SMR. Upon crossing the N\'e{}el temperature into the antiferromagnetic phase of \ch{Cr2O3}, the SMR signal decreases by more than one order of magnitude. Thus, the SMR can be used to probe the magnetic phase transition of the AFM in thin film AFMI/Pt microstructures.

We thank H. Gomonay, J. Barker and U. K. R{\"o\ss}ler for fruitful discussions as well as M. Lammel and S. Piontek for technical support. We acknowledge financial support by the Deutsche Forschungsgemeinschaft via SPP 1538 (project no.\ GO 944/4 and TH 1399/5).

\bibliography{Cr2O3_SMR}

\end{document}